\documentclass{aa}  

\usepackage{graphicx}
\usepackage{color}
\usepackage{booktabs}
\usepackage{verbatim}
\usepackage{txfonts}
\usepackage{tablefootnote}
\usepackage[dvipsnames]{xcolor}

\makeatletter

\newcommand{\Rmnum}[1]{\expandafter\@slowromancap\romannumeral #1@}

\newcommand{\NI}{N\Rmnum{1}}
\makeatother

\begin{document}

   \title{Detection of Nitrogen gas in the $\beta$ Pictoris circumstellar disk}


   \author{P. A. Wilson\inst{1,2,3,4}
          \and
          R. Kerr\inst{5}
          \and
          A. Lecavelier des Etangs\inst{3,4}
          \and
          V. Bourrier\inst{3,4,6}
          \and\\
          A. Vidal-Madjar\inst{3,4}
          \and
          F. Kiefer\inst{3,4}
          \and
          I. A. G. Snellen\inst{2}
          }

   \institute{Department of Physics, University of Warwick, Coventry CV4 7AL, UK\\
\email{paul.a.wilson@warwick.ac.uk}
\and
   Leiden Observatory, Leiden University, Postbus 9513, 2300 RA Leiden, The Netherlands
   \and
   CNRS, UMR 7095, Institut d'Astrophysique de Paris, F-75014, Paris, France
         \and
         Institut d'Astrophysique de Paris, F-75014, Paris, France
         \and
             University of British Columbia, 2329 West Mall, Vancouver, BC, V6T 1Z4, Canada
        \and
            Observatoire de l'Universit\'e de Gen\`eve, 51 chemin des Maillettes, 1290 Sauverny, Switzerland
             }

  \date{Received August 29, 2018; accepted November 9, 2018}
  
  \abstract
  {The debris disk surrounding $\beta$ Pictoris has a gas composition rich in carbon and oxygen, relative to solar abundances. Two possible scenarios have been proposed to explain this enrichment. The preferential production scenario suggests that the gas produced may be naturally rich in carbon and oxygen, while the alternative preferential depletion scenario states that the enrichment has evolved to the current state from a gas with solar-like abundances. In the latter case, the radiation pressure from the star expels the gas outwards, leaving behind species less sensitive to stellar radiation such as C and O. Nitrogen is also not sensitive to radiation pressure due to its low oscillator strength, which would make it also overabundant under the preferential depletion scenario. As such, the abundance of nitrogen in the disk may provide clues to why C and O are overabundant.}
  {We aim to measure the nitrogen column density in the direction of $\beta$ Pictoris (including contributions by the interstellar medium and circumstellar disk), and use this information to disentangle these different scenarios explaining the C and O overabundance.}
  {Using far-UV spectroscopic data collected by the {\emph{HST}}'s \textit{Cosmic Origins Spectrograph} (COS) instrument, we analyse the spectrum and characterise the {\NI} triplet by modelling the absorption lines.} 
  {We measure the nitrogen column density in the direction of $\beta$ Pictoris for the first time, and find it to be $\log(N_{\mathrm{N\Rmnum{1}}}/1\,\mathrm{cm}^2) = 14.9\pm0.7$. The nitrogen gas is found to be consistent with solar abundances and Halley dust.  We also measure an upper limit for the column density of Mn\Rmnum{2} in the disk at $\log(N_{\mathrm{MnII}}/1\,\mathrm{cm}^2)_{\mathrm{CS}}=12.7^{+0.1}$ and calculate the column density of S\Rmnum{3}$^{**}$ in the disk to be $\log(N_{\mathrm{SIII}^{**}}/1\,\mathrm{cm}^2)_{\mathrm{CS}_{X}} = 14.2\pm0.1$. Both results are in good agreement with previous studies.}
  {The solar nitrogen abundance supports the preferential production hypothesis, in which the composition of gas in $\beta$\,Pictoris is the result of photodesorption from icy grains rich in C and O or collisional vaporisation of C and O rich dust in the disk. It does not support the hypothesis that C and O are overabundant due to the insensitivity of C and O to radiation pressure thereby leaving them to accumulate in the disk.}


   \maketitle
%

\section{Introduction}
\label{sec:introduction}
$\beta$ Pictoris ($\beta$ Pic) is a young  $23\pm3$\,Myr \citep{mamajek_2014} planetary system which hosts a near edge-on debris disk composed of dust and gas. The gas in the disk is not thought to be primordial, but rather continually replenished through evaporation and collisions between dust grains. The detection of the CO molecule \citep{vidal-madjar_1994, dent_2014, matra_2017}, is evidence that new gas is being produced. This is because of the very short, $\sim120$\ year typical lifetime of the CO molecule \citep{visser2009}. The presence of UV photons from the ambient interstellar medium rapidly dissociates CO into C and O, which then evolves by viscous spreading \citep{kral_2016, kral_2017}.

Observations have shown that the circumstellar (CS) disk has a particularly large overabundance of C and O relative to a solar abundance \citep{roberge_2006, brandeker_2011}. \cite{xie_2013} introduced two opposing hypotheses for explaining the origin of this overabundance. The first hypothesis, which they named {\it preferential production}, occurs when the gas is produced with an enriched C and O abundance. Such an enrichment could be created through processes which release gas from solid bodies with an inherently high C and O abundance, such as bodies rich in CO \citep{kral_2016}. The main gas producing processes are photodesorption off the grains; a non-thermal desorption mechanism which releases gas by UV-flux \citep{grigorieva_2007}; collisional vaporisation of the dust in the disk \citep{czechowski_2007}; cometary collisions \citep{zuckerman_2012}; and sublimating exocomets \citep{ferlet_1987, beust_1990, Lecavelier96}. {\it Preferential depletion}, the second of the two hypotheses by \cite{xie_2013}, suggests that the gas evolves to have an elevated C and O abundance from an original solar abundance ($\gtrsim 1$ dex). Metallic elements subject to strong radiation forces, such as Na and Fe, could deplete more quickly than C and O for which the radiation force is negligible, which would lead to an overabundance of C and O. The presence of the overabundant C and O has been invoked as an explanation for why other gas species seen in the CS disk have not been blown away by the radiation pressure of the star and why the orbital motion of the gas is consistent with Keplerian rotation \citep{fernandez_2006}.\

A column density measurement of nitrogen could help disentangle these two hypotheses for explaining the overabundance of C and O. If the N/Fe column density is found to comparable to N/Fe solar abundance, this will speak in favour of {\it preferential production}. If the column density of N is found to be high, this could support the {\it preferential depletion} hypothesis. This is because N, like C and O, is not sensitive to radiation forces and thus should under the {\it preferential depletion} scenario be overabundant compared to radiation-sensitive species.

A column density measurement of N might also shed light on the gas production mechanism in the $\beta$ Pic CS disk. Molecular abundance analyses of comets in the Solar System indicate that cometary coma consist mostly of H$_2$O ($\sim90\%$) followed by CO ($\sim5\%$) and CO$_2$ $\sim3\%$ and that Nitrogen-bearing molecules such as N$_2$, NH$_3$, and CN  are only minor constituents (e.g. \citealt{krankowsky_1986,eberhardt_1987,wyckoff_1991}). Although these studies have measured the cometary coma and not the internal abundances, it is unlikely that comets should produce significant amounts of N.

In this paper, we present the first detection of nitrogen in the $\beta$ Pic disk based upon observations obtained using the Hubble Space Telescope ({\it{HST}}) together with the far-UV Cosmic Origins Spectrograph (COS) instrument. In Sect.\,\ref{sec:observations} we present the observational setup and technique used. The analysis of the data, which includes airglow removal, spectral alignment, and the combination and modelling of the {\NI} lines is presented in Sect.\,\ref{sec:analysis}. The results are presented in Sect.\,\ref{sec:results} with a discussion of the obtained {\NI} abundance measurements in Sect.\,\ref{sec:discussion}. The main findings and significance of the {\NI} abundance measurement is summarised in Sect.\,\ref{sec:conclusions}.


\section{Observations}
\label{sec:observations}
The far-UV observations of $\beta$\,Pic were obtained using the {\it{Cosmic Origins Spectrograph (COS)}} on the {\it{Hubble Space Telescope (HST)}} using the TIME-TAG mode and the G130M grating. The observations were done using the primary science aperture, which has a 2.5\,{\arcsec} diameter field stop. The observations consisted of a total of 11 visits and 31 orbits, conducted between February 2014 and May 2018. The central wavelength was first set at 1291\,\AA\, however, upon inspection of the data, a decision was made to change the central wavelength setting to 1327\,\AA\ for all observations after 2016 to avoid the photon poor region towards the shortest wavelengths. The two central wavelength positions overlap in the 1171 - 1279\,\AA\ and 1327 - 1433\,\AA\ wavelength regions.

To avoid contamination by geocoronal emission (airglow) we use the Airglow Virtual Motion (AVM) technique where the target is deliberately offset along the dispersion axis thereby separating the target spectrum from the stationary airglow spectrum. A more detailed description method of this method can be found in the Appendix section of \cite{wilson_2017}. The data were all reduced using the {\tt COSTOOLS} pipeline \citep{fox15} version 3.2.1 (2017-04-28).

\section{Analysis}
\label{sec:analysis}

\subsection{Airglow modelling and subtraction} \label{sec:agm}

The Hubble Space Telescope is located in Earth's tenuous atmosphere (in the exosphere) where H, N and O is present in sufficient quantities to be detected in emission at far-UV wavelengths. These emission lines, known as airglow, are present in the data, and can, if left un-corrected, reduce the accuracy of the column density estimates. We remove the airglow contamination by subtracting an airglow template made from combining several airglow-only observations. The airglow templates were created using data obtained during our previous observations \citep{wilson_2017} in combination with airglow templates made available on the Space Telescope Science Institute (STScI) website\footnote{{\url{http://www.stsci.edu/hst/cos/calibration/airglow.html}}}.

The final airglow-subtracted data, $F$, were created by subtracting the airglow template, $F_{\mathrm{AG}}$, scaled by a factor $C$, from the original contaminated data spectrum, $F_{\mathrm{tot}}$. This is expressed in the following equation:
\begin{equation}
\label{eqmod}
F = F_{\mathrm{tot}} - C \times F_{\mathrm{AG}}.
\end{equation}
\noindent A detailed explanation of these latter steps, including the methods for calculating $C$, are explained in the Appendix sections \ref{sec:AG_creation} - \ref{sec:AG_verification}.

\subsection{Choice of line spread function}
Mid-frequency errors caused by polishing irregularities in the HST primary and secondary mirrors causes the spectroscopic line-spread function (LSF) to exhibit extended wings with a core that is broader and shallower when compared to a Gaussian LSF. For our modelling we chose to use the tabulated LSF (LP3) available on the STScI website\footnote{\url{http://www.stsci.edu/hst/cos/performance/spectral_resolution/}}. Since the shape of the LSF changes with wavelength we chose the tabulated LSF closest to our wavelength region of interest: G130M/1222. The shape of the LSF reduces our ability to detect faint and narrow spectral features. It is therefore likely that we are not particularly sensitive to the fainter Mn\Rmnum{2} and S\Rmnum{3}$^{**}$ which exist close to the {\NI} lines. Furthermore, the non-Gaussian line wings also mak it hard to detect closely spaced narrow spectral features. This limits the complexity of our model to only a few absorption components.

\subsection{Modelling the N\Rmnum{1} lines}
\label{sec:modelling}
The {\NI} absorption lines in the $\beta$\,Pic spectrum were modelled using Voigt profiles. We compute the Voigt profile (a convolution of a Gaussian and a Lorentzian profile) as the real part of the Faddeeva function, computed with standard Python libraries ({\tt{scipy.special.wofz}}). Each nitrogen line in the main triplet (1199.5496, 1200.2233 and 1200.7098 {\AA}) was modelled and fit simultaneously over a predefined spectral region (1198.48 - 1202.35\,\AA). The additional {\NI} triplet at (1134.1653, 1134.4149 and 1134.9803 {\AA}) towards the end of the detector and the weak lines at 1159.8168 and 1160.9366\,{\AA} have weak line strengths which results in line depths which lie below the local noise level, so they were not included in the fitting. Once we found a satisfactory model for the main {\NI} triplet at $\sim1200$\,\AA\, we compare this solution with the other weaker {\NI} line regions and find that our model is consistent with these much weaker lines. We also modelled the Mn\Rmnum{2} and S\Rmnum{3}$^{**}$ lines at 1199.39, 1201.12\,\AA\ and 1200.97\,\AA\ respectively as they partially overlap with the {\NI} lines. The separate S\Rmnum{3}$^{**}$ line at 1201.73\,\AA\ was used to break any degeneracy between the {\NI} line at 1200.71\,\AA\ and the S\Rmnum{3}$^{**}$ line at 1200.97\,\AA.

\subsubsection{Choosing the number of model components}
We initially fit the absorption lines using a two component model consisting of one component representing the interstellar medium (ISM) and another the circumstellar gas disk. The resulting fit was poor, suggesting the presence of a third component. This third component is added and named CS$_\mathrm{X}$, and its inclusion decreased the value of the reduced $\chi^2$ from 1.6 (with 10 free parameters) to 1.0 (with 15 free parameters). This decreased the Bayesian information criterion \citep{schwarz1978} from 646 to 451 and the Akaike information criterion \citep{akaike1974new} from 607 to 409. Adding a fourth component, however, did not improve the fit, so we settle on this three-component model. The CS$_\mathrm{X}$ component, which was free to vary in radial velocity space, gave the best fit close to the rest frame of $\beta$\,Pic at 20.0\,km/s \citep{Gontcharov06, brandeker04}. The continuum was modelled using a second order polynomial.

\subsubsection{Selecting the free component parameters}
In the absence of reliable lines for which to check the wavelength calibration locally, we take the liberty of not constraining the ISM ($v_{\mathrm{ISM}}$) and CS$_0$ ($v_{\mathrm{CS_0}}$) bulk velocities to their literature values. However, since the ISM component has been well-constrained in previous studies to be offset by -10\,km/s relative to the star's radial velocity component \citep{vidal-madjar_1994, Lallement95}, we set this requirement and allow the two parameters to vary together, provided that $v_{\mathrm{CS_0}} - v_{\mathrm{ISM}} = 10$\,km/s is maintained. Aside from that constraint, all parameters for the ISM absorption component were kept fixed except for the ISM nitrogen column density, $\log(N_{\mathrm{N\Rmnum{1}}}/1\,\mathrm{cm}^2)_{\mathrm{ISM}}$. The fixed ISM parameters were $T_{\mathrm{ISM}} = 7000$\,K (the temperature of the ISM) and $\xi_{\mathrm{ISM}}=1.5$\,km/s (the turbulent broadening of the ISM), both values of which follow from studies observing the interstellar medium towards $\beta$ Pic and other nearby stars, such as \citet{Bertin95, Lallement95}.

For both circumstellar components, the column density and turbulent broadening ($\xi$) were set as free parameters. The bulk velocity of the CS$_{\mathrm{X}}$ component was also allowed to vary freely. The temperature of the CS gas is unknown. However, to avoid the degeneracy introduced when modelling both the turbulent velocity and the temperature simultaneously we fix the CS gas temperature to 10\,K for both CS components. The relationship between the parameters are expressed mathematically as
 
\begin{equation}
b^2 = \frac{2kT}{m} + \xi^2
\end{equation}

\noindent where $b$ is the width of the lines, $T$ the temperature, $k$, the Boltzmann constant, $m$ the mass of the considered species and $\xi$ the turbulent broadening (also sometimes referred to as microturbulent velocity).

\subsection{Fit to the data}
We fit the {\NI} triplet in conjunction with the Mn\Rmnum{2} and S\Rmnum{3}$^{**}$ lines with Mn\Rmnum{2} overlapping the blue side of bluest {\NI} line. We do not model S\Rmnum{3}$^{**}$ in the ISM component as sulphur will instead exist in the form of S\Rmnum{2}. We checked and found that only a model with a low S\Rmnum{3}$^{**}$ column density provides a fit consistent with the data.

\begin{figure*}
   \centering
   \includegraphics[width=\hsize]{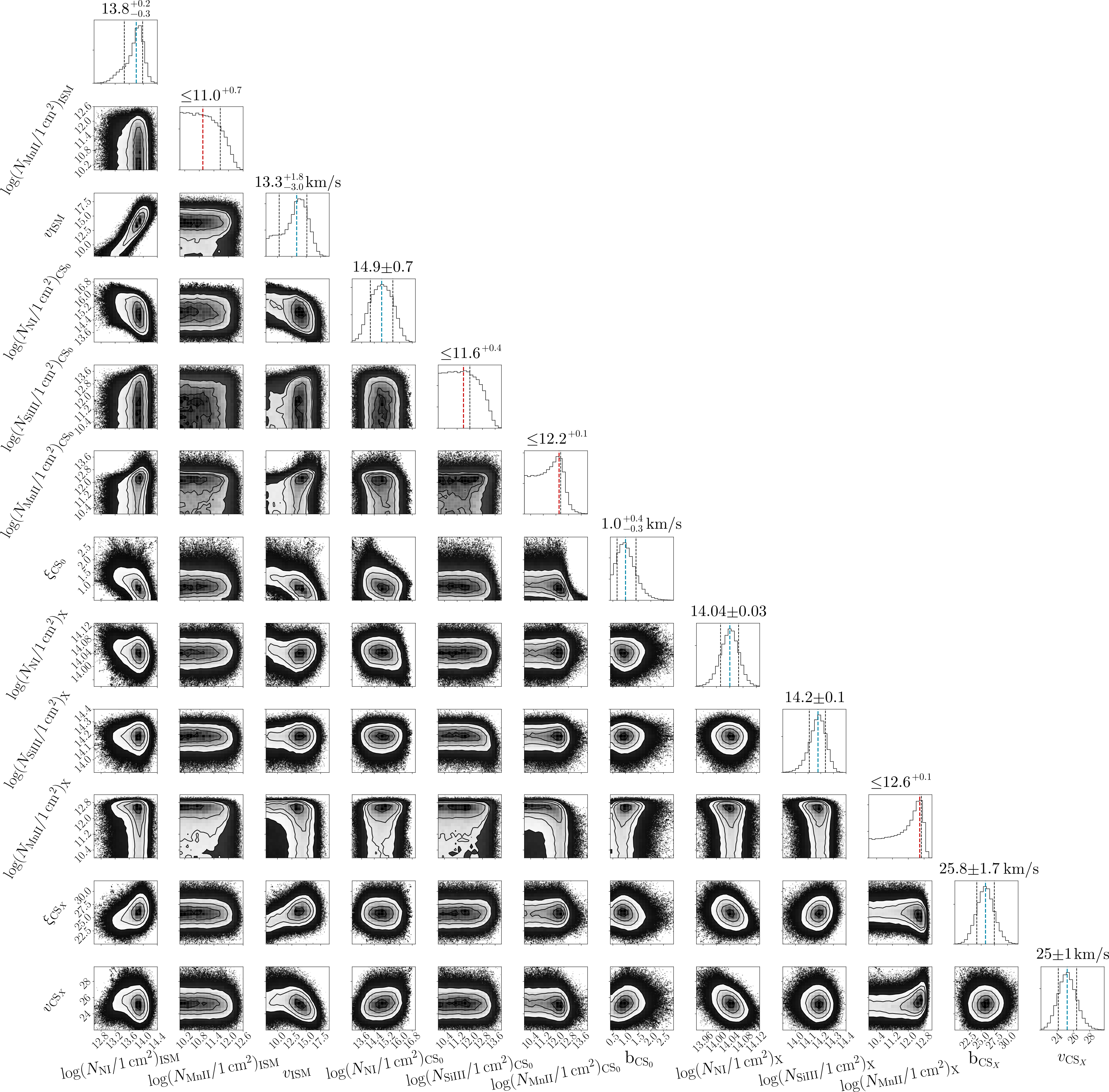}
      \caption{A corner plot showing the one and two dimensional projections of the posterior probability distributions for the free parameters. The blue dashed lines indicate the mean of the distribution. The black dashed lines in each of the 1D histograms represent the 1\,$\sigma$ deviations (68\,\% of the mass) with the central dashed line indicating the median value. The solid black lines in each of the 2D histograms represent the 1, 2 and 3\,$\sigma$ levels respectively (with 1\,$\sigma = 39.3$\,\% of the volume). The red dashed lines indicate the mode of the distribution. For the parameters which only have an upper uncertainty, the uncertainty was calculated using assuming a truncated Gaussian distribution.
              }
         \label{fig:mcmc_tabulated}
   \end{figure*}

We employed a least squares optimisation to the data using the {\tt{scipy.optimize.leastsq}} package \citep{scipy} to obtain initial starting parameters for a Markov chain Monte Carlo (MCMC) run using the {\tt emcee} code described in \citet{dfm_2013}. We ran 200 MCMC walkers with a total of 10,100 steps each, with 100 burn in steps. This resulted in a total of two million steps, of which $\sim25$\,\% were accepted. The posterior probability distributions for the free model parameters are shown in the corner plot \citep{dfm_2016} in Fig.\,\ref{fig:mcmc_tabulated}, along with the marginalised 1D distributions which were used to calculate the median values and the uncertainties on each free parameter (see Table\,\ref{table:params}).

\section{Results}
\label{sec:results}

\begin{table}
 \caption[]{The medium values and uncertainties of the posterior probability distributions generating using the MCMC method.}
\label{table:params}
\begin{tabular}{lcc}
 \hline \hline
\noalign{\smallskip}
Parameter & Value\\
\hline

\noalign{\smallskip}
\noalign{\smallskip}
\multicolumn{2}{c}{Absorption by the ISM} \\
\noalign{\smallskip}
\hline
\noalign{\smallskip}
$\log(N_{\mathrm{NI}}/1\,\mathrm{cm}^2)_{\mathrm{ISM}}$         & $13.8^{+0.2}_{-0.3}$\\
\noalign{\smallskip}
$\log(N_{\mathrm{MnII}}/1\,\mathrm{cm}^2)_{\mathrm{ISM}}$       & $\leq11.0^{+0.7}$      \\
\noalign{\smallskip}
$T_{\mathrm{ISM}}$        &      7000\,K$^\dagger$\\
\noalign{\smallskip} 
$b_{\mathrm{ISM}}$&                   1.5\,km/s$^\dagger$\\
\noalign{\smallskip}
$v_{\mathrm{ISM}}$               & $13.3^{+1.8}_{-3.0}$\,km/s\\
\noalign{\smallskip}
\hline

\noalign{\smallskip}
\noalign{\smallskip}
\multicolumn{2}{c}{Narrow absorption by gas at $\beta$\,Pic system velocity, CS$_0$} \\
\noalign{\smallskip}
\hline
\noalign{\smallskip}
$\log(N_{\mathrm{NI}}/1\,\mathrm{cm}^2)_{\mathrm{CS}_0}$     &      $14.9\pm0.7$ \\
\noalign{\smallskip}
$\log(N_{\mathrm{SIII}^{**}}/1\,\mathrm{cm}^2)_{\mathrm{CS}_0}$   &    $\leq11.6^{+0.4}$    \\
\noalign{\smallskip}
$\log(N_{\mathrm{MnII}}/1\,\mathrm{cm}^2)_{\mathrm{CS}_0}$   &      $\leq12.2^{+0.1}$      \\
\noalign{\smallskip}
$b_{\mathrm{CS}_0}$                                     &   $1.0^{+0.4}_{-0.3}$\,km/s\\
\noalign{\smallskip}
$v_{\mathrm{CS}_0}$           &    $v_{\mathrm{ISM}}+10$\,km/s\\
\noalign{\smallskip}
\hline

\noalign{\smallskip}
\noalign{\smallskip}
\multicolumn{2}{c}{Broad absorption by gas at $\beta$\,Pic system velocity, CS$_{X}$} \\
\noalign{\smallskip}
\hline
\noalign{\smallskip}
$\log(N_{\mathrm{NI}}/1\,\mathrm{cm}^2)_{\mathrm{CS}_{X}}$     & $14.04\pm0.03$\\
\noalign{\smallskip}
$\log(N_{\mathrm{SIII}^{**}}/1\,\mathrm{cm}^2)_{\mathrm{CS}_{X}}$   &  $14.2\pm0.1$    \\
\noalign{\smallskip}
$\log(N_{\mathrm{MnII}}/1\,\mathrm{cm}^2)_{\mathrm{CS}_{X}}$   &   $\leq12.6^{+0.1}$    \\
\noalign{\smallskip}
$b_{\mathrm{CS}_{X}}$                                         & $25.8\pm1.7$\,km/s\\
\noalign{\smallskip}
$v_{\mathrm{CS}_{X}}$                                      & $25\pm1$\,km/s  \\
\noalign{\smallskip}
\hline
$^\dagger$ indicates the fixed parameters.
\end{tabular}

\end{table}

   \begin{figure*}
   \centering
   \includegraphics[width=\hsize]{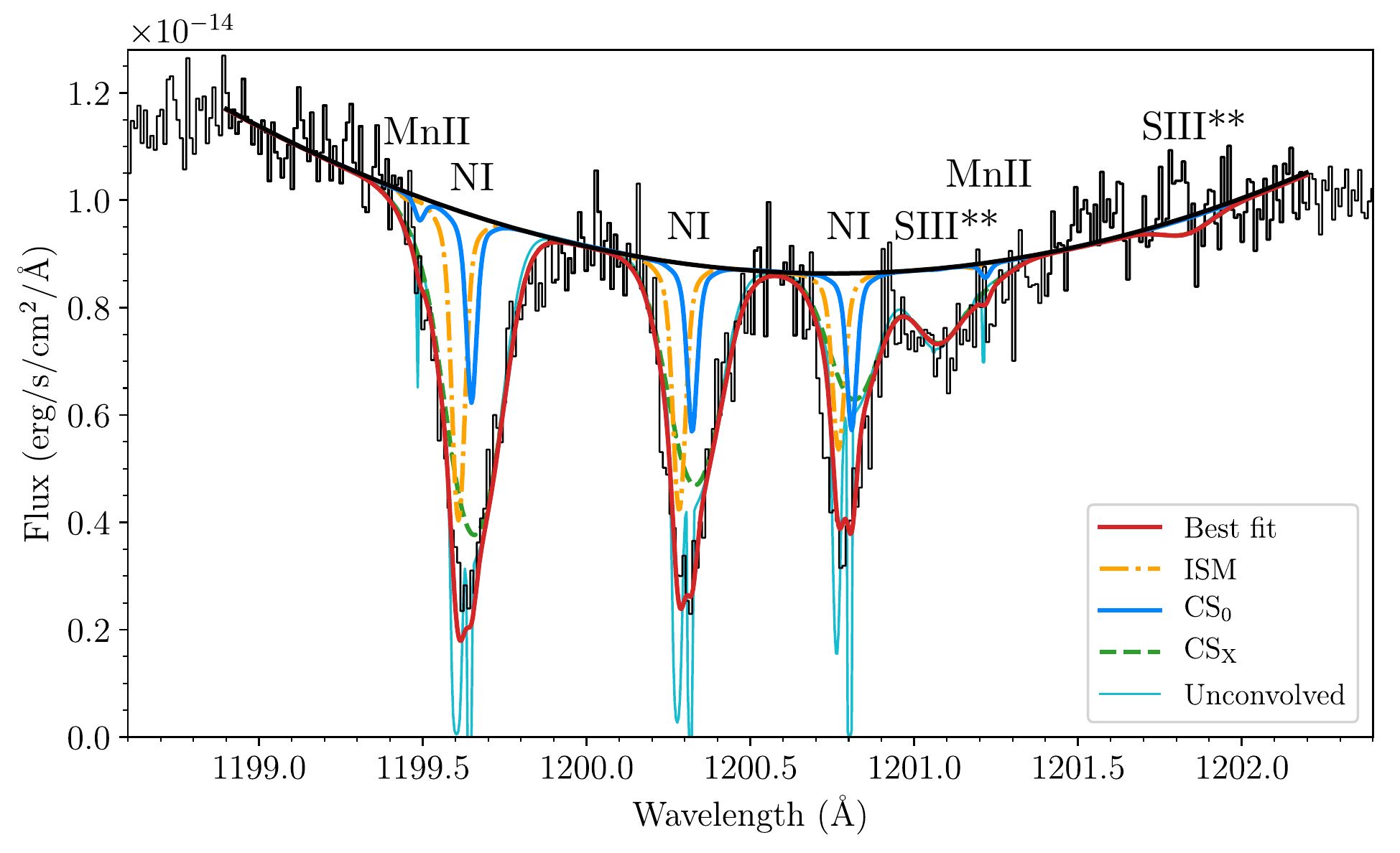}
      \caption{Measured flux as a function of wavelength (HST rest frame) for the {\NI} lines at $\sim1200$\,\AA. The black spectrum represents the combined data with the thicker parts indicating the region used to fit the continuum which is shown as a black solid line. The coloured lines show the individual components with the cyan coloured line showing the individual absorption profiles before they get convolved by the instrumental LSF.}
         \label{fig:NI_triplet}
   \end{figure*}

  \begin{figure}
   \centering
   \includegraphics[width=\hsize]{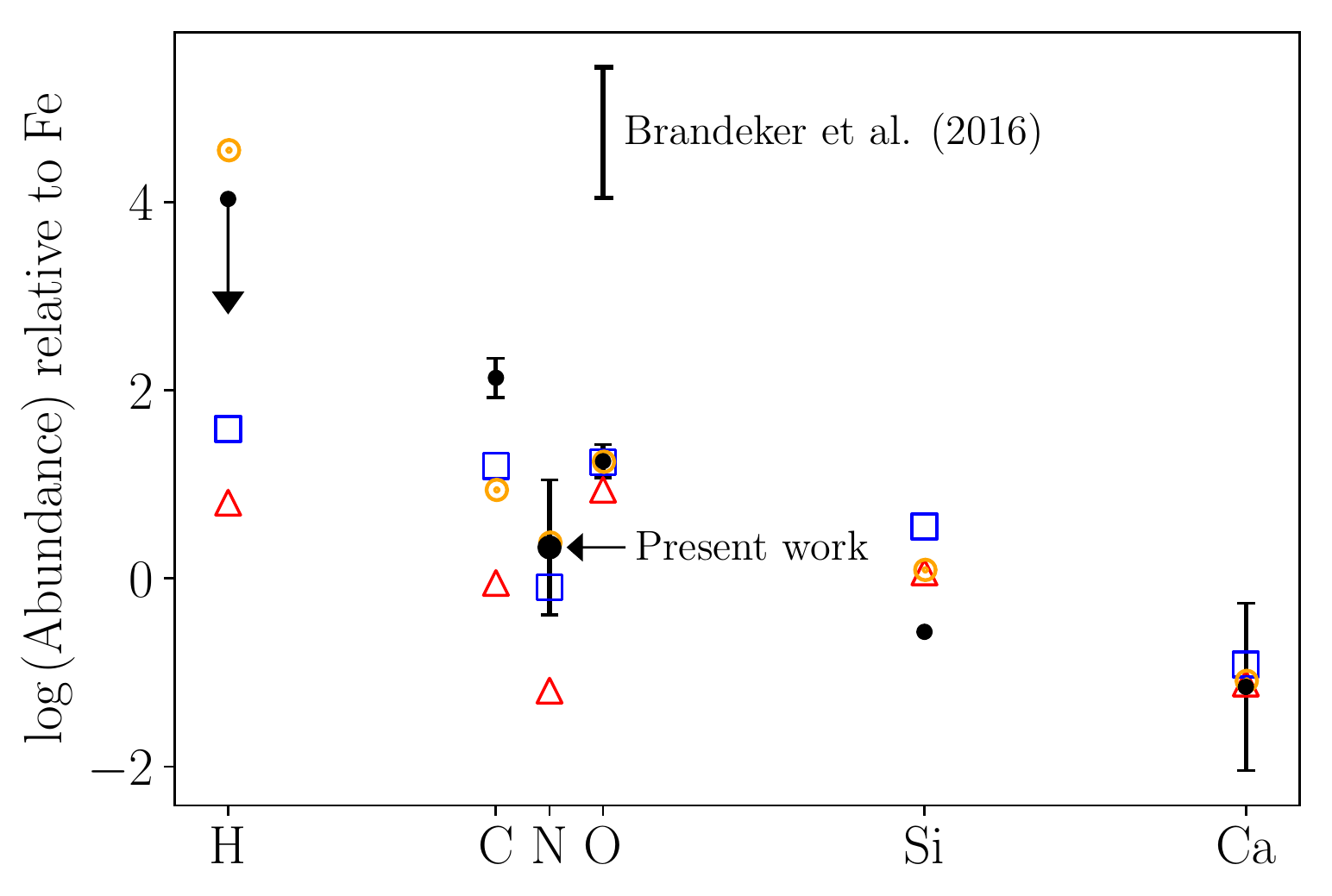}
      \caption{The abundances of the $\beta$\,Pic gas disk (black circles) compared to solar abundances (orange Sun symbols), CI chondrites (red triangles) and Halley dust (blue squares). The abundances are given relative to iron (Fe) and the figure is adapted from Fig. 2 from \citep{roberge_2006}. The O\Rmnum{1} detection by \citealt{brandeker_2016} who used the Herschel telescope, represents a range of possible values which depend on the spatial distribution of the O gas in the CS disk. The O\Rmnum{1} column density estimate by \citealt{roberge_2006} (represented by the black circle) could represent a lower column density limit due to challenges associated with measuring optically thick O\Rmnum{1} lines (see \citealt{brandeker_2011} and Fig. S1 in \citealt{roberge_2006}). The CI chondrites consistently show low abundances for the volatile elements H, C, N, O which is expected as they were in a gaseous state when the meteorite formed and did thus not condense or accrete.
              }
         \label{fig:abundances}
   \end{figure}

\subsection{Saturated N\Rmnum{1} lines}
\label{sec:saturated_lines}
The results from the best fit (and MCMC) show that the CS$_0$ component is completely saturated (seen by the cyan coloured lines which reach a flux of 0 in Fig.\,\ref{fig:NI_triplet}). The ISM component is partially saturated. The saturation leads to larger uncertainties on $\log(N_{\mathrm{NI}}/1\,\mathrm{cm}^2)_{\mathrm{CS}_0}$ reaching a magnitude uncertainty of 0.7. This is because an increase in column density doesn't make the absorption signature much deeper, but instead widens the absorption profile which is then partially absorbed by the CS$_{X}$ component. This is seen by the broad posterior distribution of $\xi_{\mathrm{CS}_{X}}$ values shown in Fig.\,\ref{fig:mcmc_tabulated}. Interestingly, although the CS$_{X}$ is free to vary in radial velocity it has the same radial velocity as the CS$_{0}$ component (within $1\sigma$). The same bulk velocity values indicates that the CS$_{0}$ and CS$_{X}$ could be describing the same gas. This is discussed more in the discussion (\S\,\ref{sec:discussion}).

We find a system velocity for $\beta$\,Pic of $23.3^{+1.8}_{-3.2}$\,km/s which is consistent with accepted value of $\sim20.5$\,km/s \citep{hobbs85,brandeker04}. The slight difference between the values are likely caused by uncertainties in the wavelength solution.

To estimate the total column density of circumstellar {\NI} gas we add the two circumstellar components together:

\begin{equation}
\label{eq:col_add}
\begin{split}
\log(N_{\mathrm{\NI}}/1\,\mathrm{cm}^2)_{\mathrm{CS}}
&= \log \left(10^{\mathrm{\NI}_{\mathrm{CS}_0}/1\,\mathrm{cm}^2} + 10^{\mathrm{\NI}_{\mathrm{CS}_{X}}/1\,\mathrm{cm}^2} \right)\\
&= 14.9\pm0.7
\end{split}
\end{equation}

\noindent We compare the ratio of the N/Fe column density (normalised to Si) with solar abundances, CI chondrites and Halley dust in Fig.\,\ref{fig:abundances}. It shows that the {\NI} abundance is consistent with solar and Halley dust abundances, yet is inconsistent with CI chondrites.

\subsection{An indirect estimate of the H column density of the ISM in the direction of $\beta$\,Pic}
Using the {\it HST} and the Goddard High Resolution Spectrograph (GHRS), \cite{meyer97} conducted observations of the interstellar {\NI} and found a value for the mean interstellar gas-phase N/H abundance at $(7.5\pm0.4)\times10^{-5}$. Furthermore, they reported no statistically significant variations in the measured N abundances from sight line to sight line and no evidence of density-dependent nitrogen depletion from the gas phase.
We estimate the total hydrogen column density in the ISM given the value for the {\NI} ISM column density calculated in this paper and assuming N({\NI})$ \approx $N({N$_{\mathrm{total}}$}). From the {\NI} ISM column density we calculate that the H\Rmnum{1} ISM value in the direction of $\beta$\,Pic is $\log(N_{\mathrm{HI}}/1\,\mathrm{cm}^2)_{\mathrm{ISM}} = 17.9^{+0.2}_{-0.3}$. This is consistent with \cite{wilson_2017} where this value was found to be $18.2\pm0.1$.

\subsection{Column density estimates for Mn\Rmnum{2} and S\Rmnum{3}$^{**}$}
We are able to place robust upper limits on the column density estimates for Mn\Rmnum{2} and S\Rmnum{3}$^{**}$. The lower limit is generally unobtainable as the absorption signature is so small that it becomes indistinguishable from the noise. The Mn\Rmnum{2} posterior distributions are not flat topped, however, which one would expect when the absorbing component becomes indistinguishable from the noise. As there seems to be a preferred column density value, we estimate the Mn\Rmnum{2} column density by calculating the mode of the Mn\Rmnum{2} column density distributions and find them to be $12.2^{+0.1}$ and $12.6^{+0.1}$ for the CS$_0$ and CS$_\mathrm{X}$ components respectively. Added together we get a value of $\log(N_{\mathrm{MnII}}/1\,\mathrm{cm}^2)_{\mathrm{CS}}=12.7\pm0.1$ or $(5.0\pm1.2)\times10^{12}$\,atoms cm$^{-2}$. This value is consistent with the value of $3.8\times10^{12}$\,atoms cm$^{-2}$ reported by \cite{lagrange_1998}.

The total S\Rmnum{3}$^{**}$ CS column density is dominated by the CS$_\mathrm{X}$ component which is $\log(N_{\mathrm{SIII}^{**}}/1\,\mathrm{cm}^2)_{\mathrm{CS}_{X}} = 14.2\pm0.2$ or $(1.6\pm0.7)\times10^{14}$\,atoms cm$^{-2}$. \cite{lagrange_1998} measure a value of $1.1\times10^{14}$\,atoms cm$^{-2}$ which is also consistent with our measurement.

\section{Discussion}
\label{sec:discussion}
\subsection{Implications of a solar {\NI} abundance}

The column density of {\NI} relative to iron is shown in Fig.\,\ref{fig:abundances}. The {\NI} column density is consistent with solar abundances, unlike C and O, which are overabundant.

The {\it preferential depletion} scenario presented in \cite{xie_2013} suggest the overabundance of C and O is due to their accumulation in the disk. As the radiation from $\beta$\,Pic depletes the metallic elements (susceptible to radiation pressure), such as Na, Mg, Ca and Fe, the radiation resilient species remain in the gas disk resulting in an C and O overabundance. Under such a scenario one would expect N to also be overabundant, as N has a low sensitivity to radiation pressure, like C and O. This can be demonstrated by evaluating the ratio of the radiation pressure to stellar gravity, which is expressed as $\beta$:

\begin{equation} \label{eqm}
\beta = \frac{d^2}{G m M_\star}\frac{1}{4\pi\epsilon_0}\frac{\pi e^2}{m_e c^2}f\phi_\nu
\end{equation}

\noindent where $d$ is the distance to the star, $G$ is the gravitational constant, $m$ is the mass of the ion under consideration, $M_\star$ is the stellar mass, $\epsilon_0$ the permittivity of free space, $e$ the elementary charge, $m_e$ the mass of the electron, $c$ the speed of light, $f$ the oscillator strength of the transition and $\phi_\nu$ the stellar flux at the considered wavelength (per unit frequency). We calculate this quantity using the following values: $f=0.130$ (for the strongest line), $m=14$\,u, $\lambda=1199.55$\,\AA, stellar flux $= 1.0\times10^{-14}$\,erg/s/cm$^2$/\AA, $d=19.3$\,pc, and  $M_\star=1.75M_\odot$. This returns $\beta\simeq$ $3.6\times10^{-4}$. While determining the exact influence of radiation pressure on nitrogen atoms would require a dynamical model, such a low value for $\beta$ indicates that it should have a negligible effect. Despite these minimal effects from radiation pressure, we do not find an overabundance of nitrogen in the $\beta$ Pictoris disk. Other papers have also shown no clear deficiency in radiation sensitive species such as Na and Ca \citep{lagrange_1998}, which also seems to disfavour the preferential depletion scenario.

The {\it preferential production} scenario, also proposed by \cite{xie_2013}, suggests that the overabundance is the result of gas being produced from materials rich in C and O. This includes photodesorption from C and O-rich icy grains \citep{grigorieva_2007} and collisional vaporisation of the dust in the disk \citep{czechowski_2007}. Unlike preferential depletion, this scenario does not infer high N abundances and is thus more compatible with the near-solar {\NI} abundance measured here.

\subsection{The origin and dynamics of the {\NI} gas}
The absorption components which we use to model the shape of the {\NI} absorption lines may provide us with hints at both the origin and dynamics of the {\NI} gas. The two absorbing circumstellar components, $\mathrm{CS}_0$ and $\mathrm{CS}_X$, were not constrained to particular radial velocity values, yet provided the best fit when their radial velocities matched that of the star. This indicates an absence of detectable amounts of {\NI} falling in towards the star. The clearly asymmetric H\Rmnum{1} Ly-$\alpha$ line presented in \citet{wilson_2017} on the other hand, indicates that HI gas is falling in towards the star. The H\Rmnum{1} gas is thought to be originating from the dissociation of water molecules from evaporating exocomets. The fact that we do not see similar asymmetric line profiles in {\NI} is perhaps not surprising. In Solar System comets, nitrogen-bearing molecules are only minor constituents \citep{eberhardt_1987}, so if $\beta$ Pic comets are similar, we would not expect the {\NI} lines to be asymmetric. This does, however, not exclude the possibility that {\NI} might originate from exocomets, as we argue in \S\ref{sec:exocomets}.

\subsection{The role of exocomets}
\label{sec:exocomets}
The {\NI} column density measurements are dominated by the narrow absorption feature, CS$_0$, which has a line width of $b_{\mathrm{CS}_0}\sim1$\,km/s. With a similar line width observed in Fe\Rmnum{1} \citep{vidal-madjar_2017}, this absorption feature likely traces a stable band of circumstellar material orbiting $\beta$\,Pic. The origin of the broad absorption feature, CS$_X$, with $b_{\mathrm{CS}_X}\sim26$\,km/s which absorbs $7\times$ less but clearly governs much of the absorption line shape is less clear. It may be the case that the feature consists of a number of individual absorption components which are outside the resolving powers of the COS instrument.

The broad component could be related to {\NI} absorption lines in the stellar atmosphere of $\beta$\,Pic. This would certainly explain the bulk velocity matching that of the star. However, we would then expect the lines to be much broader and shallower due to the rapid rotation of $\beta$\,Pic at $\sim130$\,km/s \citep{royer_2007}. In any case, a slight contamination of the line by absorption in the stellar atmosphere would not alter significantly the final {\NI} column density and the subsequent conclusion.

Another possibility is that the component is caused by gas which originated from exocomets and which has accumulated over time. Although exocomets are unlikely to be rich in nitrogen, there may have been a build up of {\NI} from successive cometary visits over time. The {\NI} which does originate from comets is not going to be very sensitive to radiation pressure, and thus may linger for longer in the system. This could explain the stability of the feature and the width of the line resulting from the velocity distribution of the evaporating comets which over time deposit the {\NI}.

The exocomets in the $\beta$ Pic disk can be categorised into two separate populations with different dynamical properties (Kiefer et al. 2014). The first population (population S) consists of exocomets that produce shallow absorption lines at high radial velocities ($\sim40$ km/s and above), which are attributed to exhausted exocomets trapped in a mean motion resonance with $\beta$ Pic b. The second population consists of exocomets producing deep absorption lines at low radial velocities, which could be related to the recent fragmentation of one or a few parent bodies. The comets which may contribute to the wide absorption feature seen here would originate from this latter population (population D). The velocity distribution of these comets has a FWHM of 15$\pm$6 km/s. This line width is similar to that of the broad absorption feature, suggesting that the build-up of gas in cometary debris streams from this population may explain the line's large width. Interestingly, if the exocomets fragment, the rates at which {\NI} is produced could be higher.

\section{Conclusions}
\label{sec:conclusions}
We measure the column density of neutral nitrogen in the $\beta$~Pictoris circumstellar disk for the first time, and find it to be $\log(N_{\mathrm{N\Rmnum{1}}}/1\,\mathrm{cm}^2) = 14.9\pm0.7$. Comparing the abundance ratio of {\NI} relative to Fe in the $\beta$ Pic disk we obtain a result consistent with both solar and Halley dust abundance ratios. In addition, we measure an upper limit for the Mn\Rmnum{2} column density in the disk to be $\log(N_{\mathrm{MnII}}/1\,\mathrm{cm}^2)_{\mathrm{CS}}=12.7^{+0.1}$ and measure S\Rmnum{3}$^{**}$ to be $\log(N_{\mathrm{SIII}^{**}}/1\,\mathrm{cm}^2)_{\mathrm{CS}} = 14.2\pm0.1$, which agrees well with previous estimates.

The near-solar abundance ratio of {\NI} to Fe favours the preferential production scenario which suggests the gas surrounding $\beta$\,Pic is naturally rich in C and O. We detect two distinct absorption components; a high column density and narrow component which we attribute to a stable band of circumstellar material orbiting $\beta\,Pic$ and a broader, lower column density component which we suggest could be due to the successive build up of {\NI} which originates from exocomets.

\begin{acknowledgements}
P.A.W, A.L.E and A.V-M all acknowledge the support of the French Agence Nationale de la Recherche (ANR), under program ANR-12-BS05-0012 "Exo-Atmos". P.A.W. and I.S. acknowledge support from the European Research Council under the European Unions Horizon 2020 research and innovation programme under grant agreement No. 694513. This project has been carried out in part in the frame of the National Centre for Competence in Research PlanetS supported by the Swiss National Science Foundation (SNSF), and has received funding from the European Research Council (ERC) under the European Union's Horizon 2020 research and innovation programme (project Four Aces; grant agreement No 724427). F.K. is funded by a CNES fellowship. A.L.E, A.V-M and F.K thank the CNES for financial support P.A.W would like to thank Luca Matr{\`a} and Grant Kennedy for fruitful and inspiring discussions.
\end{acknowledgements}

\bibliography{references}

\appendix
\section{Airglow Subtraction}
\subsection{Creation of an airglow-free template}
\label{sec:AG_creation}

Before subtracting the airglow, we created a reference spectrum of airglow-free data to determine the amount of airglow that had to be subtracted for each observation. The amount of airglow contamination varies as a function of HST orbital position and not all of the spectra are noticeably affected by airglow. The airglow contamination is mostly dependent on the Sun-Earth-target angle followed by the amount of terrestrial atmosphere which HST observes through (which depends on the angle between the centre of the Earth and the target). Some of the observations show relatively minor contamination by airglow, and we use them to create this reference spectrum. 

We use the airglow-dominated Ly-$\alpha$ emission feature to identify the observations with minimal airglow \citep{Bourier18}, and classify observations with a peak Ly-$\alpha$ flux of less than $1\times10^{-12}$ergs$^{-1}$cm$^{-2}$\AA$^{-1}$ as containing negligible airglow effects in the nitrogen triplet. A weighted average of spectra with negligible airglow contamination was then computed. The average of these spectra was treated as an {\it{airglow-free}} template, which we use as a reference to check the that the airglow contamination was correctly subtracted from the remaining airglow-contaminated data. The airglow-free template was smoothed by applying a linear convolution filter with a smoothing scale of 12 pix (~0.02\,\AA\, 4.3 km/s), somewhat larger than the line spread function (LSF) of the instrument, which has a FWHM of roughly 7 pix. The smoothed airglow-free template thus avoids the introduction of localised noise variations which occur on smaller pixel length scales.

\subsection{Calculating the scaling factor, $C$}
\label{sec:AG_scaling}

Once the airglow-free spectrum was generated, it was aligned to the airglow-contaminated data, as was the airglow template. Since the airglow and airglow-free spectra have different sets of visible spectral features, they had to be aligned to each airglow-contaminated spectrum using different sets of lines. The airglow template,  $F_{\mathrm{AG}}$, could not be directly aligned to the airglow around the {\NI} triplet, as the airglow emission was much weaker than the much deeper {\NI} absorption components. Using instead the airglow-dominated Lyman $\alpha$ (Ly-$\alpha$) line (16\,\AA\,away) proved to be a much more reliable alignment reference, reliably aligning the {\NI} airglow emission excess.  A second-order polynomial was fit to the top part of the Ly-$\alpha$ emission line ($\pm 120$\, km/s), which was present in all spectra. The wavelength position of the peak of the polynomial used to quantify the line position.

With the airglow template aligned to the airglow-contaminated data, the airglow-free template could be included. The cross correlation method presented in Section 3.1 of \cite{wilson_2017} was applied in the vicinity of the {\NI} lines (1199.4-1201.0 \AA) to align the airglow-free template, the airglow-contaminated spectra and airglow spectrum for each of the airglow-contaminated observation sets. The result was that after this alignment, the airglow template was aligned to the airglow in the contaminated spectrum, and the emission lines in the contaminated spectrum were aligned to the airglow-free template's emission lines. This alignment was successful for all but one, especially noisy, spectrum from the April 22, 2017 observation, which was left out of this analysis. 

With both templates properly aligned to the data, we finally apply Equation \ref{eqmod}. For each airglow-contaminated observation, $C$ was calculated by minimising the residuals between the two sides of Equation \ref{eqmod} around the N\Rmnum{1} triplet, using the averaged airglow-free model as $F$. Once $C$ was determined, $F$ was recomputed using Equation \ref{eqmod}, finally resulting in airglow-subtracted observations, which are used in the subsequent analysis.

\subsection{Airglow subtraction verification}
\label{sec:AG_verification}
We aligned the spectra taken at different epochs using the cross correlation method mentioned in Section \ref{sec:agm} \citep{wilson_2017}. We initially chose the region of 1249 to 1255 \AA, which is centred on a pair of S\Rmnum{2} lines, due to its favourable signal-to-noise ratio and the lack of transient exocometary sulphur features \citep[e.g.][]{roberge_2002, roberge_2014}. However, upon close inspection, it was found that slight radial velocity deviations existed between different lines of the same species at the same epoch (e.g. SiII, C I). We therefore conclude that wavelength solution errors exist, and that it is consequently best to align to features as close to the desired region as possible. Since the {\NI} lines themselves (1199.4-1201.0,\AA) showed minimal evidence of red or blue-shifted cometary absorption features, we decided to align to them directly, thereby avoiding any wavelength solution problems. 

All spectra were aligned and normalized to the first spectrum taken on 24 February 2014. The normalisation was done by comparing the median flux value for all of the spectra across a wavelength range redward of the N\Rmnum{1} triplet (1210.0 to 1214.0\,\AA), known to be void of airglow and strong spectral lines. No misalignment was evident upon visual analysis, so we concluded that the alignment was successful and combined all airglow-subtracted spectra through a weighted average. This combined spectrum did not show a different shape or width compared to those of the individual observations, further verifying the quality of the alignment. This produced the final spectrum that we analyse in this paper.

\end{document}